# Graphical abstract

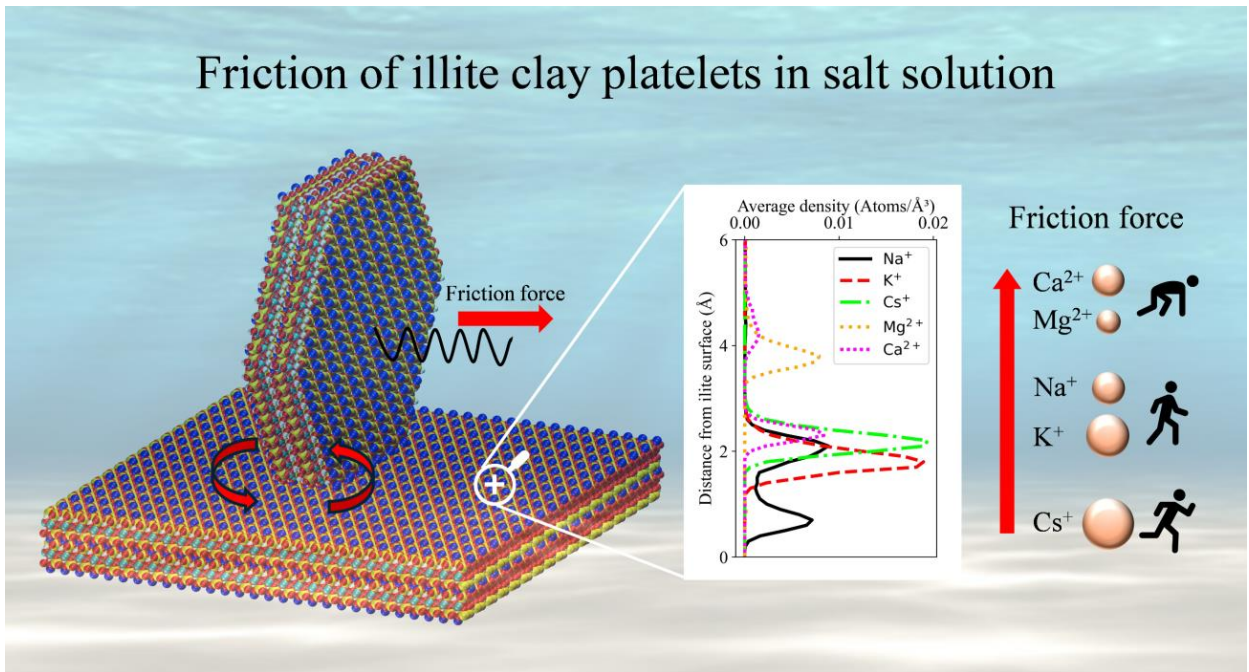

# Molecular Dynamics Simulations of Nanoscale Friction on Illite Clay: Effects of Solvent Salt Ions and Electric Double Layer

Ge Li[a,b], Astrid S. de Wijn[a,b,*]

[a] Department of Mechanical and Industrial Engineering, Norwegian University of Science and Technology(NTNU), 7491, Trondheim, Norway

[b] PoreLab, Department of Physics, Norwegian University of Science and Technology(NTNU), 7491, Trondheim, Norway

*Corresponding author: Prof. Astrid S. de Wijn

Full postal address: Richard Birkelands vei 2b, Department of Mechanical and Industrial Engineering, Norwegian University of Science and Technology(NTNU), 7491, Trondheim, Norway

Telephone: +47 92509480

E-mail address: astrid.dewijn@ntnu.no

**Abstract**

Quick clay is a highly sensitive soil that transforms rapidly from solid to liquid under minor stress, as a result of long-term salt leaching that drastically reduces shear strength. Stabilizing it is both costly and carbon-intensive, significantly impacting construction emissions in regions like Norway. Developing greener stabilization methods is challenging due to limited understanding of the weakening mechanisms and the specific roles of different salts. In this study, we use molecular dynamics (MD) simulations to investigate the sliding behavior of illite platelets, the key component in Norwegian quick clay, and how it is affected by the different ions in the solution surrounding the surface. We examine the impact of monovalent (NaCl, KCl, CsCl) and divalent (MgCl$_2$ and CaCl$_2$) salts on platelet-surface interactions, focusing on the friction enhancement brought by divalent salts and how the electric double layer (EDL) structure mediates frictional behavior. We find that divalent cations sit higher on top of the surface, and lead to an increase in friction, while monovalent cations sit closer to the surface. By providing a detailed analysis of these interactions, the study offers a novel framework for understanding the role of salts in clay mechanics and highlights opportunities to design environmentally friendly stabilizers as alternatives to traditional lime and cement.

# 1. Introduction

Quick clay is a highly sensitive soil that transitions from a solid state to a liquid form when subjected to relatively small stresses, leading to a dramatic reduction in shear strength[1,2]. This transformation poses great challenges for engineers during construction and has been the cause of numerous destructive natural landslides in high-latitude regions, including Norway (Rissa, 1978; Gjerdrum, 2020)[3,4], Sweden (Tuve, 1977)[5], and Canada (Saint-Jean-Vianney, Quebec, 1971)[6].

The concentration of salts in quick clay is critical for maintaining the structural integrity and overall strength. Quick clay originates from marine clay that was elevated above sea level and subsequently subjected to leaching by underground fresh water, which reduces its salt content. This changes the interaction between the clay particles, leading to a metastable structure often referred to as a house-of-cards structure. This house-of-cards structure is prone to collapse under external stresses, causing the release of pore water and the liquefaction of the clay. Quick clay can be made safe and stabilized using cement and lime, but this process is expensive and releases large amounts of $CO_2$[7,8]. However, the development of new and more sustainable stabilization technology is limited to trial and error approaches[9,10], due to our lack of understanding of the mechanical forces between quick clay particles on the micro and nano scale and how they are affected by the presence of salts. The goal of this study is to develop fundamental understanding of these forces, with the long-term aim of enabling development of radically new, sustainable, stabilization technology.

The primary and crucial component of Norwegian quick clay is illite, a type of phyllosilicate, along with other fractions such as chlorite, quartz, and feldspars[11]. Illite is a non-swelling, multi-layered clay with negatively charged basal surfaces and K+ ions in the interlayer that bind the layers together. When solvated in solutions with various salt types and concentrations, illite develops electric double layers (EDLs) of different structures. These EDL variations directly influence the microscopic interaction and forces between illite clay particles, thereby impacting the macroscopic strength of quick clay[12-14]. Fundamental understanding of how different salts alter the interactions between clay particles, and thus the chemical, physical, and mechanical properties of illite, is critical for understanding the mechanics of quick clay.

Illite at the microscopic scale displays a platelet-like morphology with a large lateral surface area and a small thickness. Experimentally characterizing the microscopic interactions between illite platelets and the EDL within a few angstroms outside the illite surface remains a significant challenge. Molecular dynamics (MD) simulation is a widely used approach for modeling complex structures and can be applied to mechanical properties of clay platelets at the atomic level[15]. It allows us to track all ions and analyze their effect on inter-platelet forces in detail. Previous MD simulations have successfully modeled a variety of nanoscale properties of different clays, such as the swelling and adsorption behaviors[16-21], although the majority of them have focused on infinite clay sheets. Other studies on systems containing clay platelets have examined topics like nanoscale hydration, dehydration and adsorption using atomistic models[22-24], or microscale mechanical properties using coarse-grained models[25-27]. To the best of our knowledge, no studies have yet been performed of the interactions or friction forces between the clay edge and surface using all-atom simulations that account for surface chemistry, which is of crucial importance for the forces, and thus for the objectives of this study. Nano-scale atomistic simulations of friction are computationally demanding, and very few systems of the level of complexity of clay particles have been investigated.

In this study, we investigate a crucial ingredient in the failure of the clay house-of-cards structure: the friction between the clay platelets at the nano scale. We simulate an illite platelet interacting with and sliding over an infinite illite surface. Using this model, we investigate the influences of monovalent (NaCl, KCl, and CsCl) and divalent ($MgCl_2$ and $CaCl_2$) salts on the interaction between the illite platelet and surface. Further analyses of the EDL structures near the illite clay surface provide theoretical explanations for the correlation between EDL structure and the friction changes induced by adding the above-mentioned salts. We find that the presence of divalent cations increases friction due to their location and low mobility. It is our hope that the theoretical insights gained from this research can contribute to the development of alternative stabilizers that are more sustainable than conventional options, such as lime and cement, which are associated with a high carbon footprint.

## 2. Methods

Illite clay is a multi-layered, non-swelling, dioctahedral T-O-T (tetrahedral-octahedral-tetrahedral) sandwich-like clay with the composition $KAl_2(Si_3Al)O_{10}(OH)_2$[23,28]. The initial illite structure in our simulations was derived from the unit cell of pyrophyllite-1Tc, as characterized by X-ray diffraction measurements[29]. The basic structure is shown in Figure 1. In the illite formula, 25% of $Si^{4+}$ ions in the tetrahedral layer are substituted by $Al^{3+}$ ions, resulting in a net charge of -1 per unit, which is neutralized by one $K^+$ ion. Following Löwenstein's rule, no two $Al^{3+}$ tetrahedra share corners with each other in the isomorphic charge substitutions of $Si^{4+}$ to $Al^{3+}$, i.e., preventing adjacent substitutions as shown in Figure 1(a). To balance computing resources with desirable performance, the model for simulating illite clay sliding incorporates a two-layer hexagonal illite platelet and a two-layer illite sheet in periodic boundary conditions. The latter serves as an infinite surface, enabling sufficient sliding distance for the illite platelet.

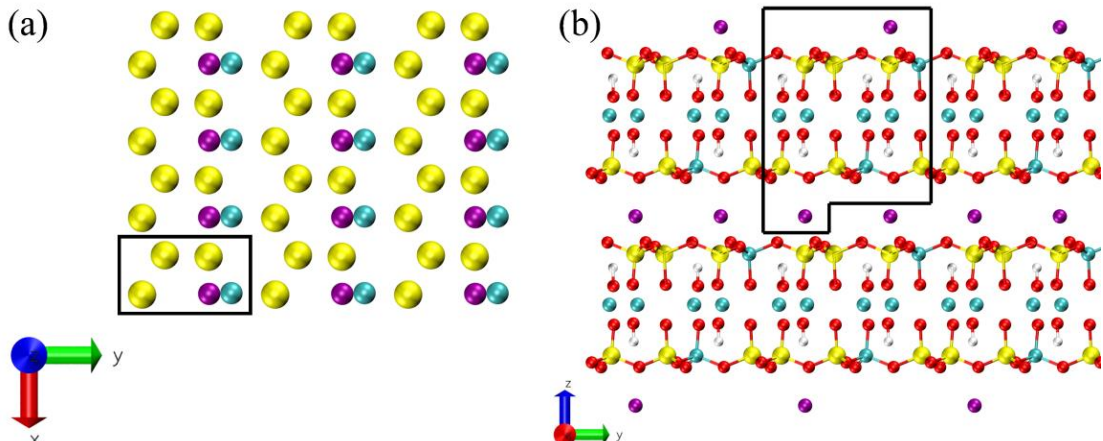

Figure 1. (a) Top view of heavy atoms only ($Al^{3+}$ in cyan, $Si^{4+}$ in yellow, $K^+$ in purple) in the partial top tetrahedral layer, indicating the isomorphic charge substitutions of $Si^{4+}$ by $Al^{3+}$ and the positions of $K^+$. The unit cell from this view is outlined with black lines. (b) a partial illite surface and the unit cell (outlined with black lines) used in this study. Color code: $O^{2-}$ red, $H^+$ white.

*2.1. Two-layer periodic illite sheet setup*

The construction of the two-layer periodic illite sheet was based on the modified 42-atom unit cell structure $K_2Al_4(Si_6Al_2)O_{20}(OH)_4$ of pyrophyllite-1Tc, with 25% $Si^{4+}$ ions replaced by $Al^{3+}$ ions. This modified unit cell shown in black lines of Figure 1(b) was subsequently replicated by 29, 17, and 2 times in the X, Y, and Z directions respectively to create a periodic illite sheet with an approximate dimensions of 15*15*2 $nm^3$. A segment of the two-layer periodic illite sheet is depicted in Figure 1(b). The negatively charged illite sheet was neutralized by placing $K^+$ ions approximately 0.23 nm above the substituted sites, with 50% of the $K^+$ ions positioned in the interlayer and the remaining 50% evenly distributed across the two outer surfaces. This distribution accounts for the presence of two interlayer surfaces and the double negative charge induced by substitutions. During a brief equilibration MD simulation, the $K^+$ ions in the interlayer become centered, while those on the outer surfaces move to their to equilibrium positions at specific distances from the surfaces. Proper $K^+$ ion distribution is particularly crucial for finite illite platelets compared to an infinite illite surface, which can be constrained and stabilized by periodic boundary conditions[30]. In contrast, an illite platelet with slanted edges and sharp corners is more prone to deformation due to inappropriate edge cleavage or uneven placement of counter ions. We discuss the effect of distribution of $K^+$ on the mechanical stability in more detail in the supplementary information. A poorly balanced distribution of $K^+$ ions can cause irreversible deformation of the one-layer illite platelet, as shown in Figure S1.

*2.2. Two-layer and four-layer hexagonal illite platelet setup*

An approximately hexagonal one-layer illite platelet was initially cleaved from a rectangular illite sheet, leading to numerous undrealistic bond cleavages along its edges with naked Al, Si or O atoms. Subsequently, this platelet was trimmed to satisfy the crystal growth theory of Hartman[31] and healed by adding hydrogen atoms or hydroxyl groups, resulting in the formation of two [010] edges and four [110] edges (Figure 2), commonly referred to as B and A/C edges, respectively, as described by White and Zelazny[32]. The detailed structures of these edges and the newly generated atoms on edges are depicted in Figure 2(b).

Two-layer illite platelets were constructed for further simulations by replicating this one-layer illite platelet along the Z dimension once, while ensuring stability by placing half of the counter ions $K^+$ in the interlayer and one-quarter $K^+$ on each surface. The two-layer hexagonal illite platelet has an edge length of approximately 5.77 nm and a thickness of 2 nm. Similarly, a four-layer hexagonal illite platelet with the same edge length but a thickness of 4 nm was built by replicating the one-layer hexagonal illite platelet along the Z direction for three times.

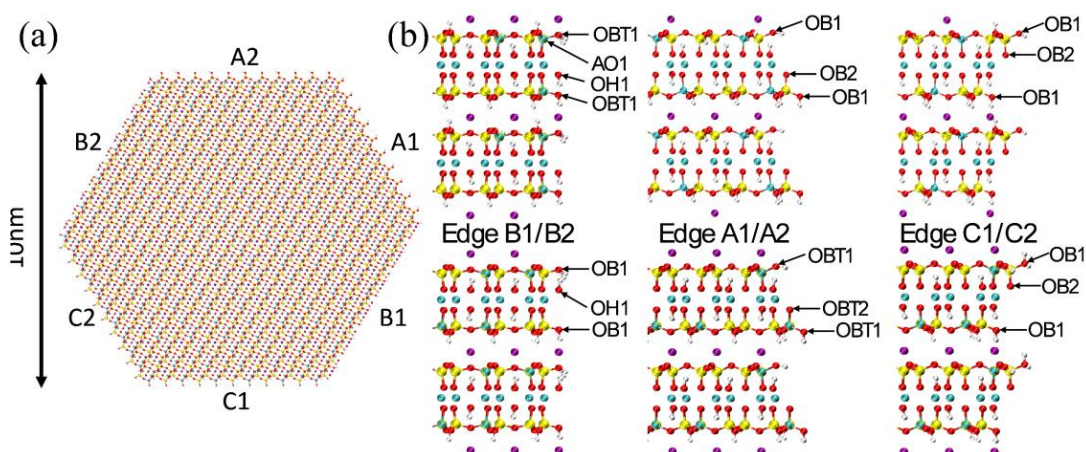

Figure 2. (a) Top view (left) of the two-layer periodic illite platelet. (b) Representation of B, A, C edges and newly generated atoms on them. Color code: $Al^{3+}$ cyan, $Si^{4+}$ yellow, $O^{2-}$ red, $K^+$ purple, $H^+$ white.

*2.3. MD simulation details*

Molecular dynamics (MD) simulations were performed using the GROMACS 2022.5 software package[33-39]. In this study, two types of systems were simulated:

System I: This system simulates illite clay edge-on-surface sliding to investigate the influence of various salts on the frictional forces of illite platelet edges against an infinite illite surface. As depicted in Figure 3, a simulation box of 15 nm * 15 nm * 20 nm was constructed to accommodate a two-layer periodic illite sheet, with the B edge of a two-layer or four-layer hexagonal illite platelet positioned perpendicular to the illite sheet. The bridging oxygen layer (OB1 and OBT1 in Figure 2) on the B edge of the illite platelet is located 6 Å away from the bridging oxygen layer of the illite surface. The flat B edge was selected to interface with the illite surface, because in our preliminary simulations we found that the A/C edges are easily damaged and prone to wearing off their hydroxyl groups even under low loads. This damage makes these edges less likely to appear in the real world and is beyond the scope of the current work. The rest of the simulation box was filled with water molecules and different salt ions and specific concentrations (NaCl, KCl, CsCl, $MgCl_2$, $CaCl_2$).

System II: In order to better understand the baseline of the illite surface and the electric double layer that the platelet in system I is sliding over, we also simulated the surface without the sliding platelet. This system includes a two-layer infinite illite surface and aimed to study the formation of the electric double layer outside the illite surface and its effect on the interaction between the illite platelet and the surface. Although the EDL thickness of illite clay is not well-documented, the very similar T-O-T clay Na-montmorillonite exhibits an EDL thickness of less than 2.5 nm[40]. In consideration of this, the construction of System II was similar to System I, except for the size of the simulation box, which was 15*15*10 $nm^3$, chosen to strike a balance between computational resources and accuracy. Our results (see section 3.3) indicate that this box size is indeed sufficient.

The initial configurations of System I and System II were prepared using the Packmol package[41], and are presented in Figure 3. The sliding model in System I comprises approximately 440,000 atoms, including 56,060 atoms representing the illite surface and platelet, while System II consists of around 220,000 atoms, with 41,412 atoms belonging to the illite sheet.

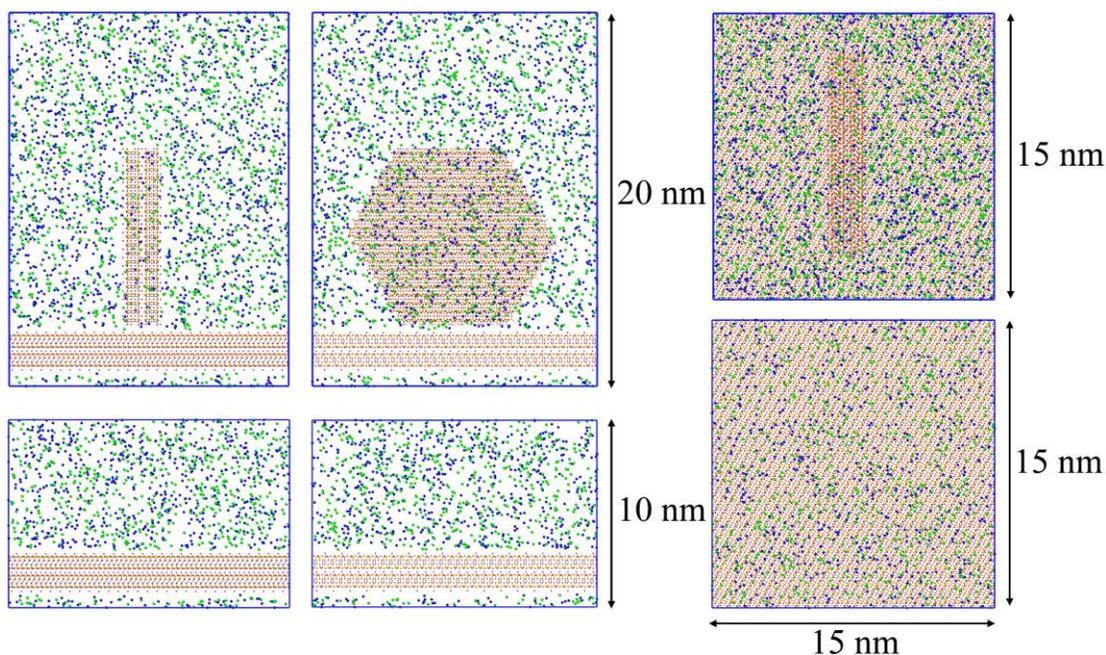

Figure 3. Front, left, and top views of System I, model for simulating illte clay sliding (the top three figures) and System II, infinite illite surface (the bottom three figures) in the NaCl-water solution. Water molecules are hidden for clarity. Color code: $Al^{3+}$ cyan, $Si^{4+}$ yellow, $O^{2-}$ red, $K^+$ purple, $H^+$ white, $Na^+$ blue, $Cl^-$ green.

Periodic boundary conditions were applied to ensure an infinite illite surface. The cutoffs for van der Waals and electrostatic interactions were set to 1.4 nm, employing the Verlet cutoff scheme[42]. Long-range electrostatic interactions beyond the cutoff were handled using the Particle-Mesh Ewald (PME) summation algorithm[43]. The O-H bonds in both water and clay were constrained using the Linear Constraint Solver (LINCS) algorithm[44]. A time step of 2 fs was used for all MD simulations. For the toms in the clay particle, we use the Clay Force Field (ClayFF)[45] and for water we use the SPC/E water model[46]. For the convenience of the reader, all relevant parameters for non-bonded and bonded interactions between original ClayFF atoms, including those from ClayFF[45] and the force field for the SPC/E water model[46] are presented in Table S1, S2 and S3. The parameters for non-bonded interactions of the newly generated atoms resulting from cleavage on B, A, C edges as shown in Figure 2 are detailed in Table 1. Their Lennard-Jones parameters were taken from ClayFF while their charges were recalculated using a bond valence model as described below[47]:

$$Z_O = -2.00 + \sum_i \frac{(Z_i - Z_i^{Partial})}{CN_i}$$

Where $Z_i$ represents the valence of the ith neighboring atom, for example, 4 for Si, 3 for Al, 1 for H. $Z_i^{Partial}$ is the particle charge of the i-th neighboring atom as defined in ClayFF. $CN_i$ denotes the coordination number of the i-th neighboring atom.

Table 1. Non-bonded parameters for newly generated atoms on edges.

| Atom name | Atom number | Mass | Charge | σ | ε | Description |
|---|---|---|---|---|---|---|
| ao1 | 13 | 26.982 | 1.8125 | 0.427124 | 5.564E-06 | Octahedral aluminum |
| obt1 | 8 | 16.000 | -1.06875 | 0.316554 | 0.650194 | Hydroxyl oxygen with tetrahedral substitution |
| obt2 | 8 | 16.000 | -1.40625 | 0.316554 | 0.650194 | Bridging oxygen with tetrahedral substitution |
| ob1 | 8 | 16.000 | -0.9500 | 0.316554 | 0.650194 | Hydroxyl oxygen |
| ob2 | 8 | 16.000 | -1.2875 | 0.316554 | 0.650194 | Bridging oxygen |
| oh1 | 8 | 16.000 | -1.1875 | 0.316554 | 0.650194 | Hydroxyl oxygen on B edges |

All MD simulations were started by performing an energy minimization simulation to eliminate unrealistic atomistic overlaps and relax the system. The relaxed system was then equilibrated for 200 ps in the NVT ensemble at 300 K using Langevin dynamics with a time constant of 1 ps for coupling[48]. For System II, a production simulation was then carried out for 30 ns at 300 K to allow counter ions, salt ions, and water molecules to fully diffuse around the illite surface and establish a stable electric double layer. Similarly, for System I, the same procedure was followed, with the addition of a 1 nN normal force applied to the illite platelet to ensure contact with the illite surface before the pulling process. For System I, after the equilibration, the pulling process starts by tethering a spring to the center of the illite platelet and dragging it along the X dimension for 20 nm to slide over the entire X length of the illite surface. Simultaneously, a normal force was applied to the platelet in the Z dimension and the angle constraints were applied to maintain the orientation of the platelet and prevent it from rotating. The pulling was conducted by dragging a spring with a force constant of 10 N/m at a rate of 1 m/s. The normal forces applied in the Z dimension were set to 0.050, 0.250, 0.500, 2.500, 5.000, 10.000, 12.500, and 25.000 nN per layer. Additionally, an extra pulling simulation was performed to test the response of the model with a thicker illite platelet. For this purpose, a four-layer illite platelet was placed on the infinite illite surface. The same range of normal forces per layer was applied on this thicker platelet. All atomistic structures presented in this study were illustrated with VMD (Visual Molecular Dynamics)[49]. The density distributions were computed in the standard way by dividing the space into slabs along the Z axis of width 0.1 Å, binning the atomic property (number, mass, charge or electron number) within each slab, and normalizing the resulting value by the slab's volume. The number density distributions of various ions presented in section 3.4 were generated using the user-friendly VMD density profile tool developed by Giorgino[50].

## 3. Results and discussion
### 3.1. Validation test of model for simulating illite clay sliding

Non-equilibrium molecular dynamics (NEMD) simulations were performed to slide the illite platelet against the illite surface under specific normal forces by pulling a spring tethered to the platelet. The forces in the spring, representing the friction forces, were recorded throughout the 20 nm or 20 ns pulling process and the friction forces per layer are shown in Figure 4(a) and 4(b).

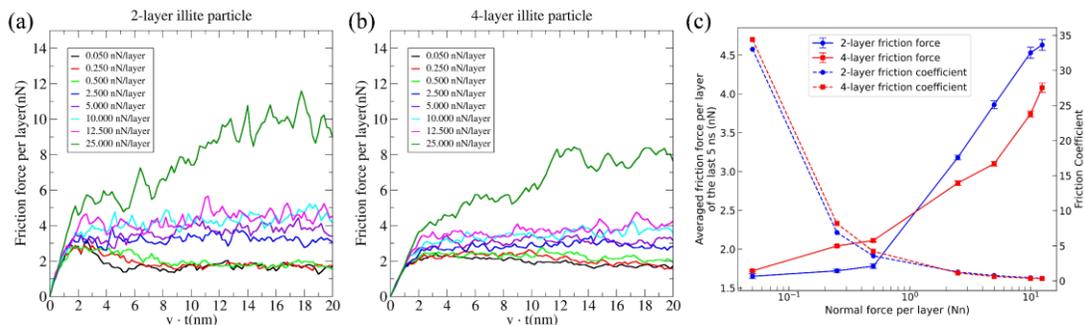

Figure 4. Friction forces per layer on two-layer (a) and four-layer (b) illite platelets under different normal forces per layer as a function of velocity · time in 0.62 M NaCl solution. The pulling velocity of the spring is 1 m/s. (c) Average friction force per layer of the last 5 nm sliding and corresponding friction coefficient vs applied normal force per layer for two-layer and four-layer illite platelets.

The friction forces generally increased with higher applied normal forces per layer and reached a plateau in time, except at extremely high load when the two-layer platelet and the four-layer platelet under a 25.000 nN/layer normal force experienced platelet breakage and subsequent folding, rendering these two force traces unreliable. We see an initial static friction that is higher than the sliding friction, especially for the two-layer illite platelet under 0.050, 0.250, 0.500 nN/layer normal forces. For higher loads, and thicker platelets, there is less distinction between static and sliding friction. In general, the thinner and thicker platelets behave in a very similar way, but there are some quantitative differences, which could be due to edge effects.

To ensure reliable analysis of the steady-state sliding, only the friction forces during the last 5 nm of the sliding process were included in our analysis of the steady state. This corresponds to the phase where the two-layer illite platelet traversed the full 15 nm distance along the X direction of the illite surface. The average friction force per layer of the last 5 ns sliding and the corresponding friction coefficient were plotted against the previously mentioned applied normal force per layer except for the largest value of 25.000 nN per layer due to the breakage and folding of the illite platelet. Results of this are shown in Figure 4(c). As the applied normal force per layer increased, the average friction force per layer during the last 5 ns of sliding also increased, while the friction coefficients dropped significantly. At low load, the friction is dominated by adhesive forces, which provide an effective load force. We observe a significant quantitative difference between the two-layer and four-layer platelets. This implies that there are finite-size effects related to the edges and surfaces of the platelets. The edges can affect the adhesive forces at low loads, while the surfaces of the platelets affect their elastic properties and may lead to the platelets deforming slightly differently under the same load per layer. At high loads, the platelets may bend in ways that significantly affect the translation symmetry.

### 3.2. Effect of different salt solutions on friction forces

As described above, salts are crucial to the interaction between illite clay platelets. In this part, various salts at a seawater concentration (approximate 0.62 M) were added to the water and the friction forces on the illite platelet were monitored as it slid across the illite surface with a total applied normal force of 10 nN on the illite platelet. This normal force of 10 nN was chosen to ensure distinct steady-state sliding with typical behavior. All the friction force traces presented in Figure S2 follow a similar trend compared to those shown in Figure 4(a) and 4(b), first increasing and then reaching a steady state. The average friction forces during the last 5 nm sliding, along with their uncertainties, are plotted in Figure S2 and indicated Table 2. The addition of all salts at a seawater concentration (approximate 0.62 M) increases the friction force, which remains the lowest in pure water for most of the last 15 nm of sliding as shown in Figure S2. The average friction force during the last 5 nm of sliding is lowest (6.28 nN) in pure water compared to all salt solutions. In Figure S2, the friction forces during the last 5 nm in $CaCl_2$ and $MgCl_2$ salt solutions are distinguishably larger than those in monovalent salt solutions. Among these, $CaCl_2$ leads to the most significant increase in friction force, reaching 9.16 nN, followed by $MgCl_2$ (8.21 nN), NaCl (7.71 nN), KCl (6.87 nN) and CsCl (6.63 nN). Our theoretical findings align with experimental observations[51], confirming that divalent $Ca^{2+}$ and $Mg^{2+}$ cations enhance the strength of quick clay more effectively than monovalent cations.

Table 2. Average friction forces and their uncertainties, estimated by the block averaging method, during the last 5 nm of sliding under a 10 nN normal force in pure water and various salt solutions at two different concentrations.

|  | Average friction forces in nN of the last 5 nm | |
|---|---|---|
|  | 0.62 M concentration | 0.31 M concentration |
| Water | 6.28±0.11 | |
| NaCl | 7.71±0.10 | 7.29±0.11 |
| KCl | 6.87±0.09 | 6.23±0.12 |
| CsCl | 6.63±0.09 | 6.07±0.10 |
| MgCl$_2$ | 8.21±0.13 | 7.65±0.13 |
| CaCl$_2$ | 9.16±0.12 | 7.93±0.09 |

The geometry of a real clay in nature is much more complex than our simulation with just a single small platelet and a single contact. In the bulk clay material, there are multiple contacts between particles and force distributions. The significant decrease in the friction force caused by removing salts like CaCl$_2$ combined with the extreme nonlinearity of this type of granular system could significantly amplify the effect on the strength in practical scenarios.

To better understand the impact of salt concentration on friction force, simulations were performed at a halved salt concentration of 0.31 M, with the corresponding average friction forces presented in Table 2. When the salt concentration was halved to 0.31 M, the friction forces decreased. However, the relative decrease is small and remained largely consistent in magnitude across different salts. The addition of CaCl$_2$, MgCl$_2$ and NaCl to water continued to enhance the friction between the illite platelet and the surface, whereas the effects of KCl and CsCl were minimal. Experimentally, the concentration of Ca$^{2+}$ and Mg$^{2+}$ in the pore water is inversely correlated with the sensitivity of quick clay, indicating that higher salt concentrations enhance its strength[52-54]. Our theoretical findings align with this trend.

*3.3. Electric double layer outside illite clay surface*

The structure of the electric double layer surrounding charged surfaces or particles plays a critical role in governing their interactions, including the friction between the illite platelet and the surface in this study. To better understand the effect of the EDL on friction, we therefore investigate it in more detail here. Figure 5(a) and 5(b) illustrates the ion density distributions near the two negatively charged illite surfaces of the infinite bi-layer sheet, located at around 11 Å and 27 Å, solvated in typical monovalent NaCl and divalent CaCl$_2$ salt solutions.

The two green peaks in each figure indicate the positions of intrinsic $K^+$ counter ions that neutralize the two negatively charged outer illite surfaces. These surface-associated $K^+$ counter ions are free to diffuse into the surrounding solution. The two $K^+$ peaks can be used as a guide for locating other ions from various salts. The middle $K^+$ peak, which corresponds to the interlayer $K^+$ counter ions that remain tightly confined between the two illite layers, is omitted to highlight the relevant range of the distribution of the other ions. The complete distributions, including the high central $K^+$ peak, along with density profiles in KCl, CsCl, and $MgCl_2$ solutions, are provided in Figure S3. Due to the symmetry of the illite surface, ions residing on one side are distributed correspondingly on the other.

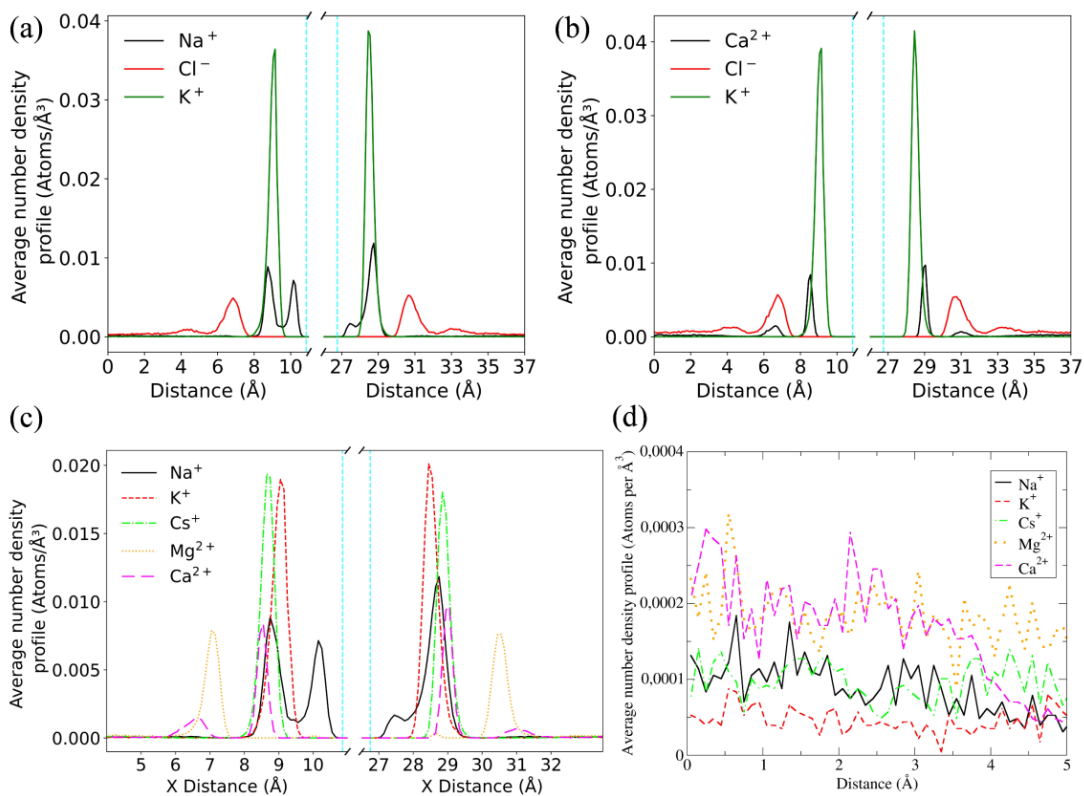

Figure 5. Density profiles of $Na^+$, $Cl^-$ and $K^+$ (a) and $Ca^{2+}$, $Cl^-$ and $K^+$ (b) outside the illite clay surface in NaCl (left) and $CaCl_2$ (right) solutions. Density profiles of $Na^+$, $K^+$, $Cs^+$, $Mg^{2+}$ and $Ca^{2+}$ (c) and the enlarged area from the bulk, 0 Å to 5 Å, (d) in different salt solutions. The two illite surfaces, defined by the outer bridging oxygen layers, are indicated by two cyan dotted lines. (a) and (b) clearly illustrate the formation of electric double layers, with most cations ($Na^+$, $K^+$, and $Ca^{2+}$) accumulating near the negatively charged illite clay surfaces, while anions ($Cl^-$) are positioned further away. The single-peaked monovalent cations, $K^+$ and $Cs^+$, exhibit more concentrated distributions, as reflected in their higher peak intensities. In contrast, the other cations ($Na^+$, $Ca^{2+}$, and possibly $Mg^{2+}$) display double peaks, with the secondary peaks of the divalent cations notably positioned further from the surface compared to those of the monovalent cations. The prominent peak for confined $K^+$ in the interlayer is omitted to enhance the visibility of other lower peaks.

As shown in Figure 5(a), apart from the main peak of the smaller monovalent

Na$^+$ located approximately 2 Å from the surface, similar to the intrinsic K$^+$ ions, a small fraction of Na$^+$ ions resides even closer at around 1 Å. In contrast, the majority of the larger divalent Ca$^{2+}$ ions are positioned slightly further at around 2.5 Å from the surface than K$^+$. Additionally, a secondary small peak of Ca$^{2+}$ appears at around 4 Å from the surface, coinciding with the main peak of Cl$^-$. A secondary small peak of Cl$^-$ is observed at around 7 Å from the surface.

The density distributions of Na$^+$, K$^+$, Cs$^+$, Mg$^{2+}$, and Ca$^{2+}$ cations are presented in Figure 5(c) when the different salts are added to the system separately. As previously described, smaller monovalent cations, like Na$^+$, reside closer to the illite surface, followed by K$^+$ and Cs$^+$. For the divalent cations Ca$^{2+}$ two distinct peaks of unequal height are observed. Mg$^{2+}$ close to the surface is not able to escape on the time scales of our simulation (see Figure 6(a) below), and we cannot therefore draw any conclusions about the height of the small Mg$^{2+}$ peaks within 3 Å from the surface. In the magnified 0-5 Å region of density profiles shown in Figure 5(d), it is evident that divalent Mg$^{2+}$ and Ca$^{2+}$ cations tend to have a small population further away from the surface than the monovalent cations. Among all cations, K$^+$ exhibits the lowest density at further distances.

The presence of these ions on the surface can have a significant impact on the friction since adsorbates at sub-monolayer coverage can impair sliding. The friction is affected strongly by the precise conditions, especially the strength of the interaction with the surface and the mobility of the adsorbate on the surface[55]. To confirm and better understand how this mechanism plays a role, we further investigate the dynamics and interactions of the monovalent and divalent cations on and near the surface.

First, we consider the time scales of the redistribution of the different cations. In Figure 6(a), 6(b) and 6(c) we show the lifetime histograms in several regions within specific distances from the illite surface during the 30 ns simulation. In the bulk, the behavior no longer changes, as indicated by the similar trends of the lifetime histograms in Figure 6(c) and far away from the surface. Close to the surface, within 3 Å of the illite surface as shown in Figure 6 (a), all three Mg$^{2+}$ stay throughout the 30 ns simulation, as can be seen from the horizontal line, and Ca$^{2+}$ migrates out slowly compared to the monovalent cations. This is likely due to the relatively small size of the divalent cations, allowing them to fit more easily between the surface atoms. In addition, they are more strongly bound due to their higher charge. The monovalent Cs$^+$ ions, which are also the largest, are completely pushed out after 3600 ps. Within 4 Å of the illite surface, shown in Figure 6(b), the lifetime histograms of K$^+$ and Cs$^+$ already become similar to the bulk ones shown in Figure 6(c). Before reaching the bulk behavior, the flatter curves of Ca$^{2+}$ lifetime histogram and the gradually constant curves of Mg$^{2+}$ lifetime histogram shown in Figure 6(a) and 6(b) suggest that divalent cations diffuse more slowly than monovalent cations near the illite surface. This finding matches the magnitude of the overall diffusion coefficients of divalent and monovalent cations in the XY dimension as discussed below.

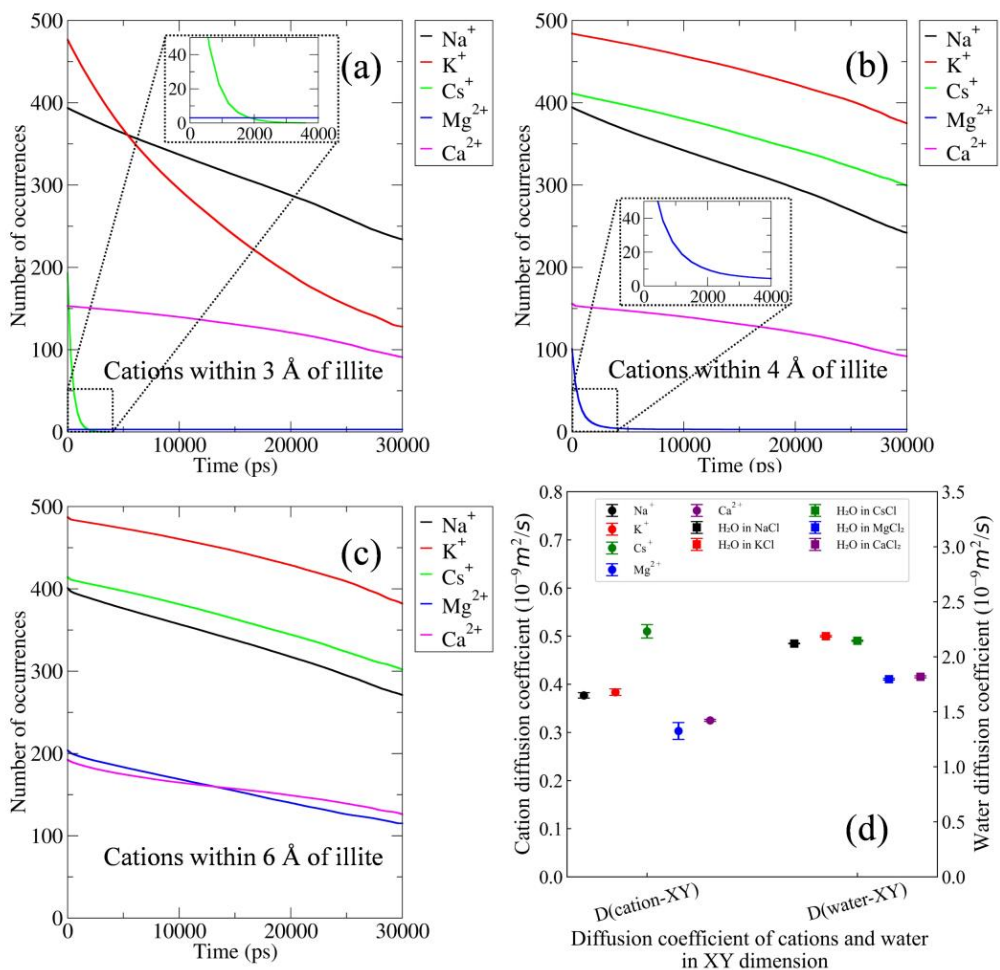

Figure 6. Lifetime histograms of $Na^+$, $K^+$, $Cs^+$, $Mg^{2+}$ and $Ca^{2+}$ within 3 Å (a), 4 Å (b) and 6 Å (c) of illite clay surface. Overall diffusion coefficients of cations and water (d) in different salt solutions in the XY dimension.

The friction is affected strongly by the mobility of cations in the XY dimension, as the sliding platelet needs to either push them out of the way or if they are not mobile enough, or move over them[55]. To probe this mobility, we investigated the overall diffusion coefficients of cations and water in various salt solutions along the XY dimension ($D_{xy}$) as shown in Figure 6(d). The less relevant overall diffusion coefficients along the Z and XYZ dimensions ($D_z$ and $D_{xyz}$) are shown in Figure S4. The overall diffusion coefficients in the XY, Z and XYZ dimensions were obtained by fitting the average mean square displacement (MSD) as a function of time over selected intervals. Specifically, a 1-10 ns window was used for cations in $D_{xy}$ while a 0.5–2.5 ns range was selected for water molecules in $D_{xy}$ based on MSD linearity (Figures S5 and S6) and computational efficiency. Divalent cations exhibit lower $D_{xy}$ than monovalent cations, with $Mg^{2+}$ showing the lowest and $Cs^+$ the highest. Water diffusion coefficients in different salt solutions follow a similar trend to those of the cations. Specifically, water diffuses more slowly in the presence of divalent salts compared to monovalent salts.

To quantify the strength between various cations and surrounding water, the non-bonded interactions between them were calculated by summing the electrostatic and Lennard-Jones potentials, yielding interaction energies of -252.45 kJ/mol for each $Na^+$, -130.89 kJ/mol for each $K^+$, -97.11 kJ/mol for each $Cs^+$, -1552.83 kJ/mol for each $Mg^{2+}$, and -1203.16 kJ/mol for each $Ca^{2+}$. These results indicate that divalent cations exhibit significantly stronger interactions with water molecules compared to monovalent cations, with $Mg^{2+}$ displaying the highest interaction energy, followed by $Ca^{2+}$, $Na^+$, $K^+$, and $Cs^+$. This trend aligns with findings from both theoretical (ab initio calculations and classical simulations) and experimental studies on cation hydration free energies[56-59], which demonstrate that smaller cations, such as $Mg^{2+}$ among divalent cations and $Na^+$ among monovalent cations, form more energetically favorable hydration shells. The higher hydration energies of divalent cations like $Mg^{2+}$ and $Ca^{2+}$ indicate the formation of robust hydration shells, where the motion of water molecules is closely coupled to the movement of the associated cations. This coupling explains the observed lower diffusion coefficients of water in solutions containing divalent salts, as the strong ion-water interactions hinder the independent movement of water molecules.

To better understand the interactions near the surface, we compute the non-bonded interactions between different parts of the illite platelet and the surrounding solution including $K^+$ counter ions, salt cations and anions, and water molecules. The results are summarized in Table 3. The illite platelet was divided into two parts, where the bottom part (≤ 6Å) resides within the Stern layer of the EDL adjacent to the infinite illite surface and the upper part (> 6Å) is immersed in the diffuse layer and bulk solution. In the Stern layer region, the bottom part of the illite platelet exhibits the strongest interaction with the $MgCl_2$-containing solution, followed by $CaCl_2$, NaCl, KCl and CsCl. For the upper part of the illite platelet, the strongest interaction occurred in the $CaCl_2$ solution, followed by NaCl, $MgCl_2$, KCl, and CsCl. This stronger interaction can hinder the sliding of the platelet, increasing friction.

The total non-bonded interactions between the entire illite platelet and the various salt solutions, as listed in Table 3, closely mirror the trend observed in the average friction forces on the illite platelet experienced during the final 5 nm of sliding.

Table 3. Non-bonded interactions in kJ/mol between different parts of the illite platelet and the surrounding solution

|  | Bottom (≤ 6Å) | Upper (> 6Å) | Total |
| --- | --- | --- | --- |
| NaCl | -11772.85 | -65847.24 | -77620.09 |
| KCl | -10774.05 | -63670.76 | -74444.81 |
| CsCl | -10745.93 | -63129.38 | -73875.31 |
| MgCl$_2$ | -14400.95 | -64327.24 | -78728.19 |
| CaCl$_2$ | -12413.12 | -67101.38 | -79514.50 |

Based on the density distributions and diffusion coefficients of various cations discussed earlier, we speculate that the increased interaction between the illite platelet and the illite surface in divalent salt solutions is directly associated with the low diffusivity and the more robust hydrated spheres of divalent cations. These cations and their associated hydrated shells diffuse more slowly and are less mobile. Furthermore, the presence of divalent cations further rather than closer to the surface hinders the movement of the illite platelet through mechanisms such as those described by Ouyang et al., 2018[55].

## 4. Conclusions

In this work, we study the mechanics of illite clay platelets at the nano scale, with the goal of developing an understanding of the mechanical behavior of quick clay. A simulation setup for investigating illite clay interactions was developed by positioning a fully flexible hexagonal illite platelet perpendicularly on an infinite illite surface. Our sliding simulations show that the addition of any salt increased the friction forces between illite clays, with divalent salts (CaCl$_2$ and MgCl$_2$) enhancing friction force more than monovalent salts (NaCl, KCl and CsCl). These molecular dynamics simulation findings align well with experimental observations of the macroscopic mechanical properties of illite clays[51].

In order to understand the mechanisms behind this different behavior and the enhanced friction, we have performed a detailed analysis of the EDL structure formed near the illite surface. Distinct peaks of divalent cations (Mg$^{2+}$ and Ca$^{2+}$), located further from the illite surface compared to those monovalent cations (Na$^+$, K$^+$ and Cs$^+$), were observed. The lower mobility of Mg$^{2+}$ and Ca$^{2+}$, combined with their strongly bound solvation shells resulting from significantly stronger hydration energies[56-59], leads to increased resistance against the sliding motion of the illite platelet. This hindering effect[55] is further supported by the stronger non-bonded interactions between the illite platelet and the surrounding solution in the presence of MgCl$_2$ and CaCl$_2$. In summary, the enhanced friction force arises from the reduced mobility of divalent cations and their hydration shells, which impede the movement of the illite platelet across the illite surface.

By studying the nanoscale friction, we are able to understand the origin of the macroscopic mechanical weakness of quick clay. The insights gained from this study contribute to a deeper understanding of how specific chemistry influences macroscopic mechanical properties, offering valuable guidance for the development of new technology and enhancing stability and performance in this highly relevant clay. Our approach using atomistic friction simulations is computationally demanding, but could otherwise easily be applied to provide similar insight into other colloidal systems.

**Declaration of Competing Interest**

The authors declare that they have no known competing financial interests or personal relationships that could have appeared to influence the work reported in this paper.

**Acknowledgement**

This work was funded by the Research Council of Norway through the project "Sustainable Stable Ground" (Project No. 324486) and its Centres of Excellence funding scheme, PoreLab (Project No. 262644). The simulations were performed using resources provided by the IDUN HPC cluster at the Norwegian University of Science and Technology (NTNU). We extend our sincere gratitude to Prof. Erika Eiser, Prof. Ida-Marie Høyvik, Dr. Seyed Ali Ghoreishian Amiri, Dr. Lu Xia, Rene Tammen, and Kamila Zablocka for their valuable input.

Supporting Information

# Molecular Dynamics Simulations of Nanoscale Friction on Illite Clay: Effects of Solvent Salt Ions and Electric Double Layer

Ge Li[a,b], Astrid S. de Wijn[a,b,*]

[a] *Department of Mechanical and Industrial Engineering, Norwegian University of Science and Technology(NTNU), 7491, Trondheim, Norway*

[b] *PoreLab, Department of Physics, Norwegian University of Science and Technology(NTNU), 7491, Trondheim, Norway*

*Corresponding author: astrid.dewijn@ntnu.no

**S1. Mechanical stability evaluation of one-layer illte platelet**

Unlike swelling T-O-T clays such as montmorillonite, where charge substitutions in the octahedral layer lead to weak surface charges and mobile counter ions, illite exhibits strong surface charges due to $Si^{4+}$ substitutions by $Al^{3+}$ in the tetrahedral layers, causing $K^+$ counter ions to remain tightly bound to the surface, which is crucial for its stability and surface properties. To evaluate the impact of ion distribution on stability, we compared the mechanical stability of two one-layer illite platelets: one with $K^+$ counter ions exclusively on one side (Platelet 1) and the other with $K^+$ ions distributed equally on both sides (Platelet 2) in vacuum and partially solvated environments. In both vacuum and 20 wt. % water conditions, Platelet 1 exhibited significant distortion (Figure S1) due to the initial asymmetric placement of $K^+$ counter ions, which remained predominantly on one side, leaving the opposite side with a strong net negative charge. This imbalance caused the positively charged side to move to cover and neutralize the negatively charged side, thereby causing irreversible deformation, manifested by the insertion of $K^+$ ions into the enlarged Si-O gaps on the upper surface of the deformed illite platelet. This deformation is consistent with the finding by Jia et al., 2023[1]. In contrast, Platelet 2 remained flat and stable throughout the entire 30 ns production run (Figure S1). This illustrates the importance of a well-balanced distribution of $K^+$ ions.

(a) 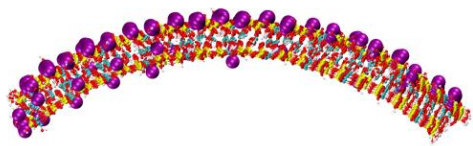  (b) 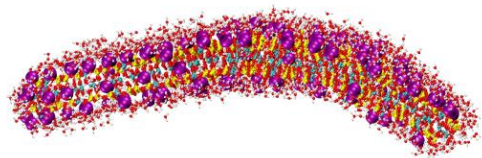

(c) 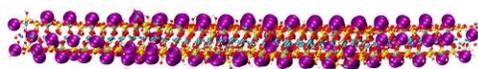  (d) 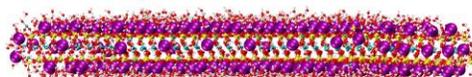

Figure S1. A one-layer illite platelet with K$^+$ on one side (Platelet 1, a and b) and a one-layer illite platelet with equivalent K$^+$ on two sides (Platelet 2, c and d) in vacuum and 20 wt. % water after 30 ns production MD simulations.

## S2. Non-bonded and bonded parameters for original ClayFF atoms

ClayFF was employed to describe both nonbonded interactions (Table S1) and bonded interactions (Table S2 and S3) among the original ClayFF atoms[2]. The parameters for nonbonded interactions of $Mg^{2+}$ are taken from Åqvist[3,4].

Table S1. Non-bonded parameters for original ClayFF atoms.

| Atom name | Atom number | Mass | Charge | s | e | Description |
|---|---|---|---|---|---|---|
| hw | 1 | 1.008 | 0.4238 | 0.0 | 0.0 | Water hydrogen |
| ho | 1 | 1.008 | 0.4250 | 0.0 | 0.0 | Hydroxyl hydrogen |
| ow | 8 | 16.000 | -0.8476 | 0.316554 | 0.650194 | Water oxygen |
| oh | 8 | 16.000 | -0.9500 | 0.316554 | 0.650194 | Hydroxyl oxygen |
| ob | 8 | 16.000 | -1.0500 | 0.316554 | 0.650194 | Bridging oxygen |
| obts | 8 | 16.000 | -1.1688 | 0.316554 | 0.650194 | Bridging oxygen with tetrahedral substitution |
| ohs | 8 | 16.000 | -1.0808 | 0.316554 | 0.650194 | Hydroxyl oxygen with substitution |
| st | 14 | 28.086 | 2.1000 | 0.330203 | 7.701E-06 | Tetrahedral silicon |
| ao | 13 | 26.982 | 1.5750 | 0.427124 | 5.564E-06 | Octahedral aluminum |
| at | 13 | 26.982 | 1.5750 | 0.330203 | 7.701E-06 | Tetrahedral aluminum |
| na | 11 | 22.990 | 1 | 0.235001 | 0.544338 | Aqueous sodium ion |
| k | 19 | 39.098 | 1 | 0.333400 | 0.418400 | Aqueous potassium ion |
| cs | 55 | 132.905 | 1 | 0.383104 | 0.418400 | Aqueous cesium ion |
| mg | 12 | 24.305 | 2 | 0.16444 | 3.663385 | Aqueous magnesium ion |
| ca | 20 | 40.078 | 2 | 0.287199 | 0.418400 | Aqueous calcium ion |
| cl | 17 | 35.453 | -1 | 0.439997 | 0.418818 | Aqueous chloride ion |

Table S2. Bond parameters for OH bonds with different types of O atoms

| Bond parameter | Bond length r in nm | Harmonic force constant $k_r$ in kj/mol |
|---|---|---|
| O-H bond in water | 0.1 | 463700 |
| O-H bond in hydroxyl group | 0.1 | 463700 |
| O-H bond with bridging oxygen with tetrahedral substitution | 0.1 | 463700 |

Table S3. Angle parameters for H-O-H and M-O-H angles

| Angle parameter | Angle q in degree | Harmonic force constant $k_q$ in kj/mol |
|---|---|---|
| H-O-H | 109.47 | 383.00 |
| M-O-H (M: metal ions) | 109.47 | 251.04 |

**S3. Friction force traces on the two-layer illite platelet in pure water and 0.62 M different salt solutions under a 10 nN normal force.**

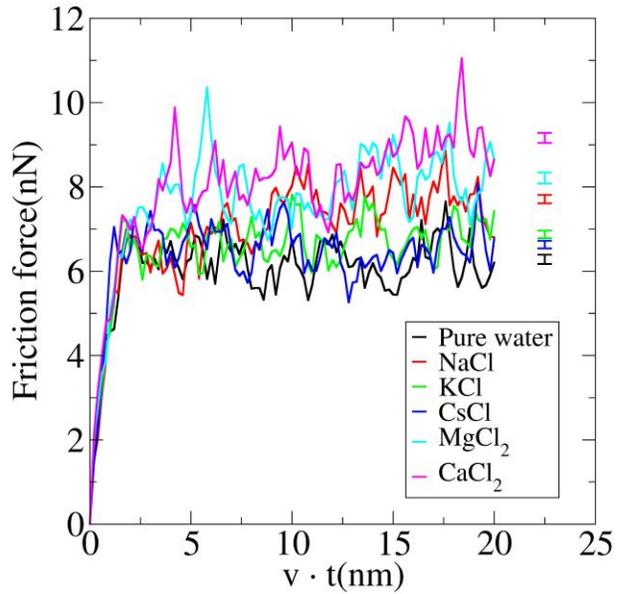

Figure S2. Friction force traces on the two-layer illite platelet in pure water and 0.62 M different salt solutions under a 10 nN normal force. The average friction forces of the last 5 nm of sliding, along with their uncertainties estimated using the block averaging method, are shown to the right of the friction force traces.

## S4. Complete density profiles of ions adjacent to the illite clay surface in NaCl, KCl, CsCl, MgCl$_2$ and CaCl$_2$ solutions

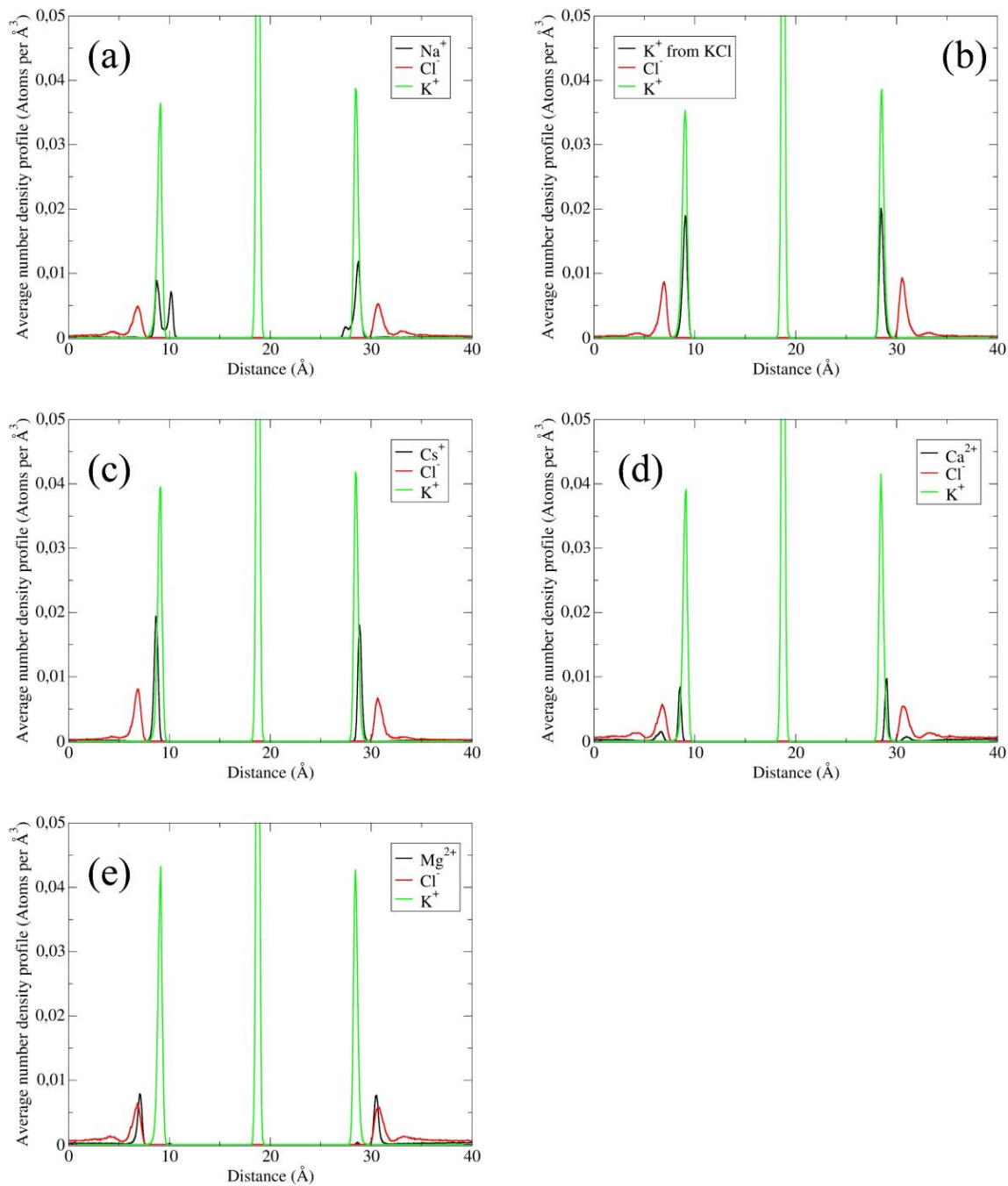

Figure S3. Density profiles of different ions outside the illite clay surface in (a) NaCl, (b) KCl, (c) CsCl, (d) MgCl$_2$ and (e) CaCl$_2$ solutions. The middle green peak, representing interlayer K$^+$, is truncated due to its excessive height, which obscures the other peaks.

## S5. Overall diffusion coefficients of cations and water in different salt solutions

The overall diffusion coefficients in the XY, Z and XYZ dimensions were obtained by fitting the average mean square displacement (MSD) in different time ranges. For cations, $D_{xy}$ and $D_{xyz}$ were fitted using average MSD data in the time range of 1 to 10 ns while $D_z$ was fitted from 0.5 to 2.5 ns, as the average MSD curve in this range exhibited better linearity. For water molecules, all diffusion coefficients were obtained from the average MSD from 0.5 to 2.5 ns due to the computational cost associated with their large number. Corresponding average MSD plots are provided in Figure S5 and S6.

The results for $D_z$ and $D_{xyz}$ are shown in Figure S4. In all cases, the average values of $D_z$ and $D_{xyz}$ are consistently smaller than the corresponding $D_{xy}$, a behavior attributed to the electric field generated by the negatively charged illite surface, which suppresses diffusion in the surface-normal direction. Moreover, the average $D_{xyz}$ of divalent cations and water in divalent salt solutions are lower than those of monovalent cations, similar to the trend observed for $D_{xy}$ as discussed in the main text. This consistent pattern highlights that the reduced diffusion of cations and hydration shells in divalent systems impedes the mobility of an illite platelet sliding across an infinite illite surface.

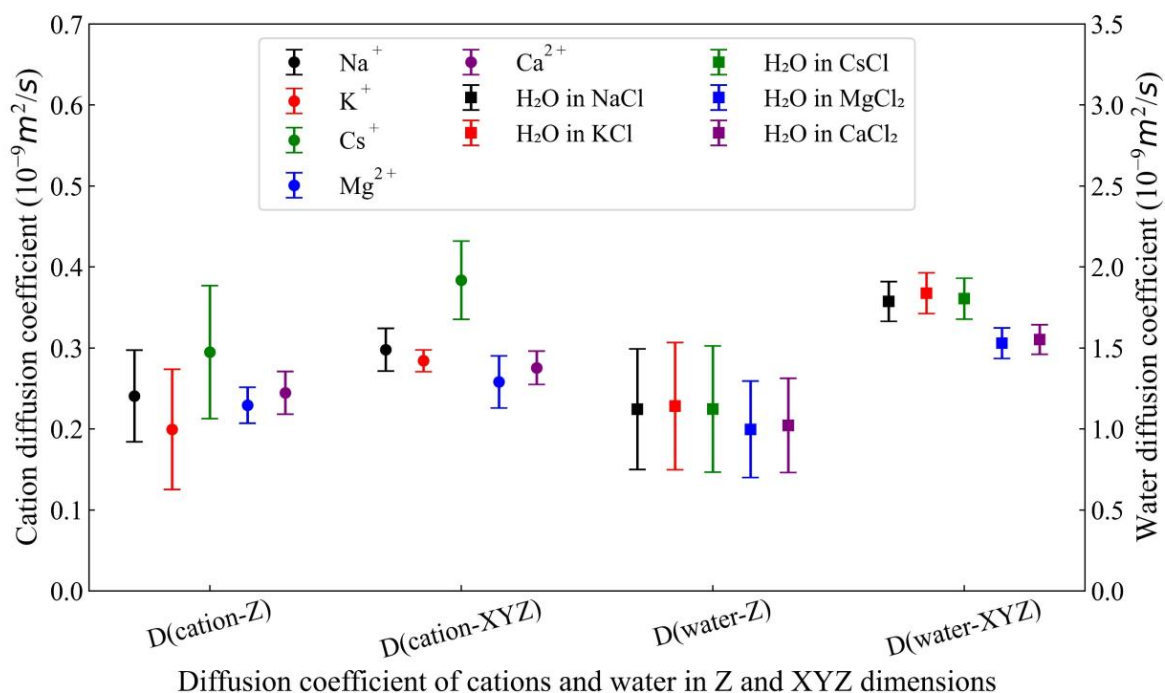

Figure S4. Overall diffusion coefficients of cations and water in different salt solutions in the Z and XYZ dimensions.

## S6. Average mean square displacement (MSD) plots of different cations in the XY, Z and XYZ dimensions over 30 ns simulations

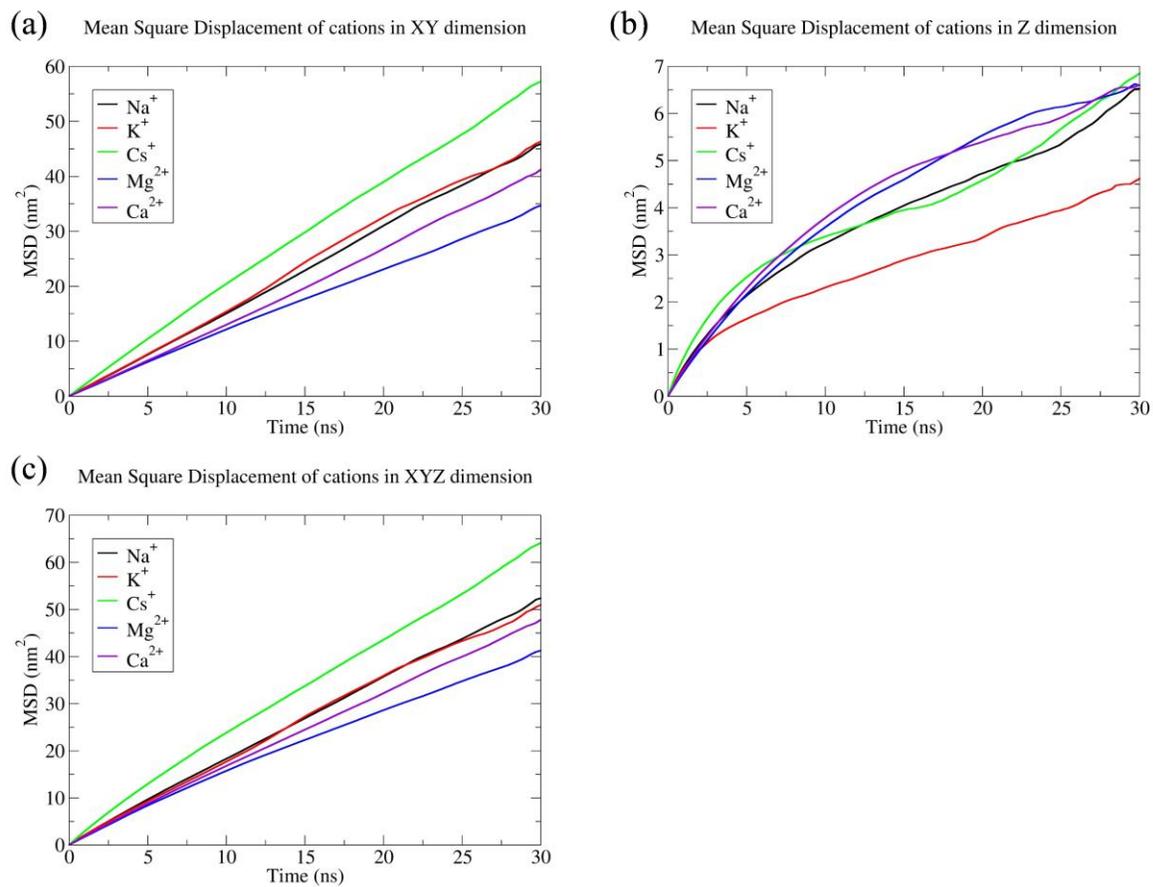

Figure S5. Mean square displacement (MSD) of $Na^+$, $K^+$, $Cs^+$ $Mg^{2+}$ and $Ca^{2+}$ cations in (a) XY, (b) Z and (c) XYZ dimensions.

## S7. Average mean square displacement (MSD) plots of water molecules in different salt-containing systems in the XY, Z and XYZ dimensions over 30 ns simulations

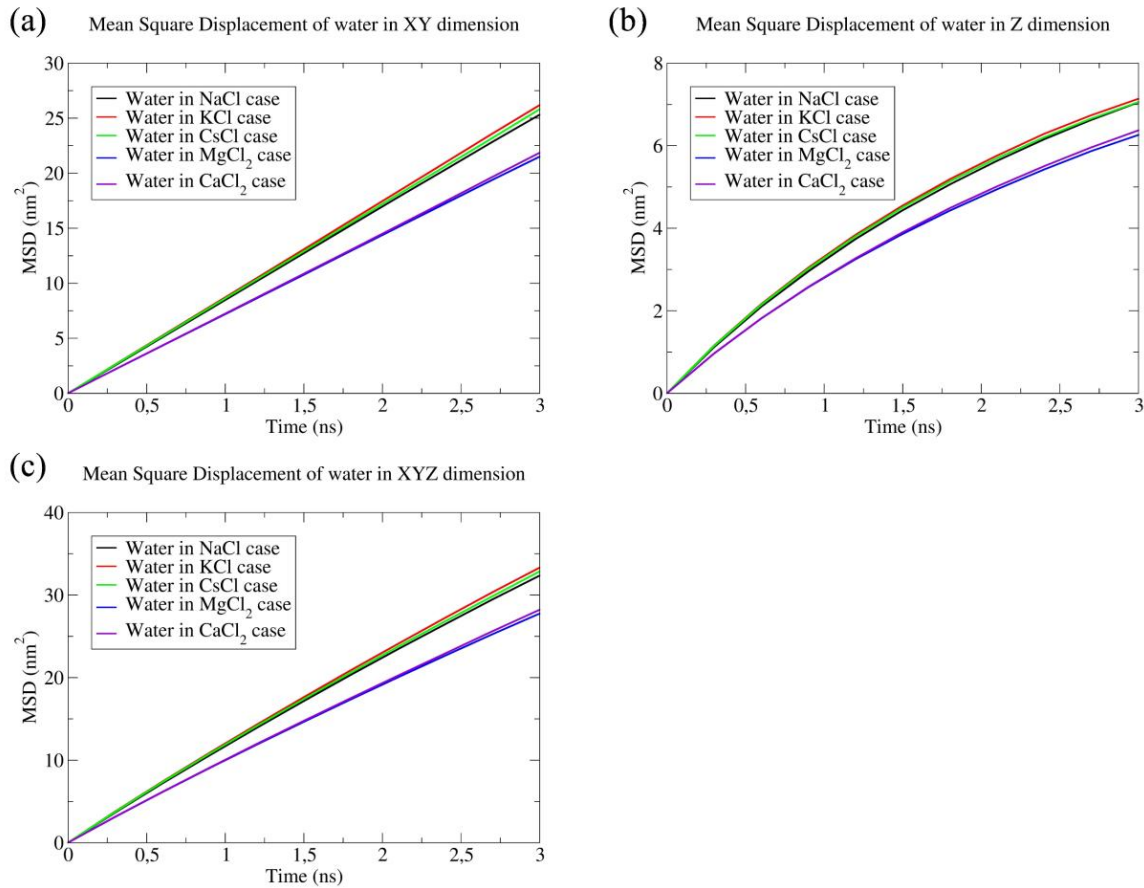

Figure S6. Mean square displacement (MSD) of water in NaCl, KCl, CsCl, $MgCl_2$ and $CaCl_2$ in (a) XY, (b) Z and (c) XYZ dimensions.